\documentclass[final,3p,times,twocolumn]{elsarticle}

\usepackage{amssymb}

\usepackage{url}

\journal{Computer Physics Communications}

\begin{document}

\begin{frontmatter}



\title{A Customized 3D GPU Poisson Solver for Free Boundary Conditions}


\author[1]{Nazim Dugan} 
\author[2]{Luigi Genovese}
\author[1]{Stefan Goedecker\corref{cor1}}
\ead{stefan.goedecker@unibas.ch}

\cortext[cor1]{Corresponding author}

\address[1]{{\it Department of Physics, University of Basel, Klingelbergstr. 82, 4056 Basel, Switzerland}}
\address[2]{{\it Laboratoire de simulation atomistique (L\_Sim), SP2M, UMR-E CEA / UJF-Grenoble 1, INAC, F-38054 Grenoble, France}}

\begin{abstract}
A 3-dimensional GPU Poisson solver is developed for all possible combinations of free and periodic 
boundary conditions (BCs) along the three directions. It is benchmarked for various grid sizes and different
BCs and a significant performance gain is observed for problems including one or more free BCs.
The GPU Poisson solver is also benchmarked against two different CPU implementations of the
same method and a significant amount of acceleration of the computation is observed with the GPU version.
\end{abstract}


\end{frontmatter}


\section{Introduction}

In the context of electronic structure calculations of molecules, wires, surfaces and solids, it is necessary to solve 
Poisson's equation with various boundary conditions (BCs), namely free BCs, wire BCs 
(periodic along one axis, free along two axes), surface BCs 
(periodic along two axes, free along one axis) and fully periodic boundary conditions.

A common method for solving the Poisson's equation 

\begin{equation}
 \label{Poisson eq}
 \nabla^2 V=\rho
\end{equation}

\noindent exploits the Fast Fourier Transform (FFT) algorithm in order to transform the input 
density $\rho$ into the Fourier space. The desired electrostatic potential $V$ can then  easily be found in Fourier space 
and it is converted back to the real
space by an inverse FFT. This FFT based solution is preferred for large data sets because of its 
$O(N \mbox{Log} N)$ scaling due to the FFTs, $N$ being the number of grid points for the discrete representation of the
density $\rho$ and the solution $V$. This method is also suitable for GPU computation since FFTs are 
efficiently implemented for GPUs by various vendors and groups \cite{Nukada.Ogata, Govindaraju.Lloyd,Volkov.Kazian,cufft}. 
Currently the NVIDIA CUFFT library supports 1D, 2D and 3D FFTs both for real and complex input data in single and double precision.

In current GPU architectures, the data array has to be 
transferred to the device memory prior to the computation and the result has to 
be transferred back to the system memory to be used in the succeeding CPU parts of the code. These data transfer times are
comparable to the GPU computation times for the Poisson solver case, reducing the possible gains of GPUs for such calculations. 
The situation is less problematic for the case of free BCs since the transferred data size
for such problems is half of the FFT size for each dimension. For a 3D input data, which is usually encountered in 
physical problems, the transferred data size is reduced to one eighth of the total FFT size, making the problem suitable 
for GPU computation. As the number of axis with free boundary conditions decreases when going from fully free boundary 
conditions to fully periodic boundary conditions, the data size increases to a quarter, a half and finally the full FFT data 
set size and the transfer times dominate more and more. Additionally, 3D FFT computation time for free BC problems can be reduced by using 
customized algorithms since initially only one eighth of the 3D FFT size is non-zero, remaining being zero padded \cite{Goedecker}. 

Some GPU based Poisson solvers can be found in the literature.
In the work of Deck and Singh \cite{Decyk.Singh} the CUFFT library is used in the FFT based solution of the 2D  Poisson's equation for periodic BCs.
Rossinelli et al. \cite{Rossinelli.Bergdorf} use the same technique for a 2D free BCs problem. However, they do not use a specialized algorithm to 
avoid calculations over zero padded regions but they apply multidimensional FFTs on the full zero padded domain. Beside these FFT based
works, there are some iterative GPU Poisson solvers \cite{Kosior.Kudela,Knittel} which do not exploit the $O(N \mbox{Log} N)$ 
scaling of FFTs.

In this work, we describe a GPU Poisson solver which uses customized 3D FFTs to increase the performance 
for free BCs. This customized GPU Poisson solver is benchmarked for various grid sizes for fully free, wire and surface BCs
to see the possible gain in the presence of free BCs. It is also benchmarked against a CPU implementation of the same method 
which uses the FFTW library \cite{Frigo.Johnson} for 1D batched FFTs. The developed GPU Poisson was implemented in the BigDFT 
wavelet based electronic structure package \cite{Genovese.Neelov, Genovese.Ospici, Genovese.Videau} and benchmarked 
against the current CPU Poisson solver of BigDFT \cite{Genovese.Deutsch} for fully free BCs.

\section{Poisson's Equation}

In the discretized integral form, the Poisson's equation (Eq. \ref{Poisson eq}) for a periodic input density becomes a convolution
of the form:
\begin{equation}
\label{convol}
 V({\bf r})= \sum_{r^{\prime}} K_p({\bf r}-{\bf r^{\prime}}) \rho({\bf r^{\prime}})~,
\end{equation}

\noindent where $K_p({\bf r}-{\bf r^{\prime}})$ is the Green's function associated with the periodic BCs. This equation 
can be solved easily in Fourier space just with a multiplication operation:

\begin{equation}
\label{convol_fourier}
V^F({\bf k})=K_p^F({\bf k}) \rho^F({\bf k})~,
\end{equation}

\noindent where the superscript $F$ denotes the Fourier transform of the corresponding quantity which can be
calculated by the FFT algorithm. Once the potential is calculated in Fourier space with 
the above formula, it can be converted back to the real space with an inverse FFT operation. The advantage of this
strategy is that the convolution in Eq. \ref{convol} is calculated in $O(N \mbox{Log} N)$ operations instead of the $O(N^2)$
operations of the direct calculation, where $N=N_x N_y N_z$ is 
the total number of grid points in the 3-dimensional input density.

Poisson's equation with free BCs can be treated similarly by using the free boundary Green's function $K_f$
instead of $K_p$ in Eq. \ref{convol_fourier}. Since the use of a FFT implies periodicity, free boundary conditions can 
only be achieved by zero padding
to the input density, doubling its size in each dimension \cite{Rossinelli.Bergdorf,Hockney.Eastwood,Qiang.Ryne}. In this way 
the wrap arounds in FFTs can be avoided
at the cost of enlarging the problem size. For a mixed BCs problem, the zero paddings should be applied for all the 
dimensions having free BCs. Examples of these mixed BCs problems are surfaces with one free dimension and
wires with two free dimensions for which the associated Green's functions have to be used in the calculation \cite{Cerioni.Genovese}. 

\section{Customized GPU Poisson Solver}

In our GPU Poisson solver, we implemented the method used in the Poisson solver of the BigDFT electronic
structure package \cite{Genovese.Deutsch}. For free BC dimensions, the input density is zero padded 
doubling the data size for that dimension. Direct application of 3D FFT libraries for such a zero 
padded input data is not efficient since these algorithms work on the whole range regardless of the data being zero or not. 
In our developments, a customized 3D FFT is used for this problem which avoids the calculation of 1D FFTs 
over the zero padded parts of the input density. 

Basically, a 3D FFT can be 
calculated in three steps each having 1D FFTs of all the rows in a separate dimension.  In our 
customized 3D FFT, zero padded parts of the data are identified for each dimension and the
unnecessary calculations over these parts are avoided. To understand the algorithm let us
denote the data size in three dimensions as $N_x$,$N_y$ and $N_z$ and the BCs as $B_x$,
$B_y$ and $B_z$ having values $1$ for periodic and $2$ for free BCs. Then the FFT size for each
dimension is given as $S_j=N_j \times B_j$. In the first step, 1D FFTs of length $S_x$ should 
be calculated for $N_y \times N_z$ rows instead of the full range of $S_y \times S_z$ rows. For the 
second dimension, 1D FFTs of length $S_y$ should be calculated for $S_x \times N_z$ rows since non-zero
elements emerge in the zero padded parts of the $x$-dimension after the first step. In the
third step the whole size of $S_x \times S_y$ rows should be processed for $S_z$ length 1D FFTs. Therefore,
for a fully free boundary problem, the total number of 1D FFT operations is reduced to $7/12$ (approximately 
$58\%$) of the direct calculation of the whole zero padded range with a 3D FFT library such as CUFFT3D.
Inverse of the zero padding operations are carried out for the inverse 3D FFT after the multiplication in Eq. 
\ref{convol_fourier}, as compactifications of the data by taking out unnecessary parts. Therefore,
the same amount of reduction in the computation time is also valid for the inverse 3D FFT. For the 
heterogeneous BCs cases (surface and wire BCs), the reduction amount of 1D FFT operations is 
influenced by the choice of the axes along which free dimensions are applied  (one axis for surface 
BCs and two axes for wire BCs). If the free dimension(s) is/are chosen as the last dimension(s) of the
forward transform then the zero paddings can be postponed and the performance of the method increases.
In the inverse transform, free BC dimension(s) are processed first since the order of axes reverses for this 
case. For the wire BCs, if the 
periodic dimension is chosen as the first dimension of the forward 3D FFT then the reduction of 1D FFT operations 
is the same as the fully free BCs discussed above. The total performance for wire BCs is slightly better
than the fully free BCs of same zero padded size since for the wire case there is no need for the initial 
zero padding operation before the FFTs in the first dimension (also the final compactification in the inverse FFT). 
For the surface BCs, by choosing the free BC dimension as the last dimension of the forward 3D
FFT, the number of the 1D FFT operations reduces to $2/3$ (approximately $66\%$) of the fully periodic case.  

In the calculation of  a 3D FFT, a transposition operation usually follows each of the three steps described above 
in order to conserve the desired access pattern for optimal performance \cite{Goedecker1}. Such transpositions are 
also used in our customized 3D GPU FFT in order to preserve the coalesced access pattern to the global memory for 
the computations in three separate dimensions.
If the zero paddings for the free BCs treatment are applied in the beginning of the 
calculation for all free boundary dimensions then the
zero parts of the data should also be processed during the transpositions. In order to avoid these
unnecessary operations, zero padding operations are applied for dimensions having free BCs just before the FFTs in that 
dimensions by a spread operation which doubles the data length for each row and sets the first half to zero.
(In the inverse FFT, compactification operations are carried out after the FFTs in each dimension having free BCs, as throwing
out the second half of each row.) 
In this strategy the transpositions are applied with reduced data sets and additionally the zero padding spread GPU kernel 
(also the compactification kernel in inverse FFT) can be merged with the preceding transpose GPU kernel (except for the first 
dimension which does not have the transposition) 
to reduce the number of {\it read/write} operations to the GPU global memory. As an example, for fully free BCs
the first transposition operation between the first two dimensions of the forward transform is carried
out with a data size which is a quarter of the total zero padded FFT size since the second and the third dimensions are 
not zero padded at that moment. The second transposition between the second and the third dimensions is carried 
out with the half of the total zero padded FFT size since the last dimension is not zero padded yet. The 
final transposition after the third dimension should be operated on the full size but this transposition can be
eliminated completely with a transposition of the associated kernel of the convolution in Eq. \ref{convol_fourier} which is 
carried out only once in the multiple application of the Poisson solver which is usually the case in realistic applications. 

Since the input density for the Poisson solver is always a real valued quantity it would be advantageous to use 
real to complex FFTs for the first dimension of the forward 3D FFT. Because of the property 
$F_{N-n}=F_n^*$ of a real to complex FFT it is enough to keep first $S_x/2+1$ terms of the result 
after the 1D FFTs in the first dimension and the succeeding 1D FFTs in the remaining dimensions are carried out over this reduced
data set of size $(S_x/2+1)\times S_y \times S_z$. It is not necessary to unfold the result at the
end of the FFT for the Poisson solver case since an inverse FFT follows the forward FFT yielding a real potential at 
the end. The complex to real 1D FFTs for the last dimension of this inverse transform has the same symmetry 
mentioned above. 

Separate steps of the customized GPU algorithm are given below where bold text is used to indicate the GPU kernels. Processed data
size is indicated for each 1D FFT step. The \textbf{\textit{transpose}} or \textbf{\textit{transpose+spread}} 
(\textbf{\textit{compactification+transpose}} for inverse FFT) kernels process same size of data as the preceding 1D FFTs:

{\scriptsize \it 
\begin{itemize} \itemsep2pt \parskip1pt \parsep1pt

\item if (B$_x$ is free) \textbf{\textit{spread\_X}}

\item \textbf{\textit{1DFFT\_X}} ($N_y \times N_z$ rows, length $S_x$) (R2C)

\item if (B$_y$ is free) \textbf{\textit{transpose+spread\_Y}}

else \textbf{\textit{transpose}}

\item \textbf{\textit{1DFFT\_Y}} ($(S_x/2+1) \times N_z$ rows, length $S_y$)

\item if (B$_z$ is free) \textbf{\textit{transpose+spread\_Z}}

else \textbf{\textit{transpose}}

\item \textbf{\textit{1DFFT\_Z}} ($(S_x/2+1) \times S_y$ rows, length $S_z$)

\end{itemize}

\begin{itemize}
\item \textbf{\textit{convolution kernel multiplication}}
\end{itemize}

\begin{itemize} \itemsep2pt \parskip1pt \parsep1pt
\item \textbf{\textit{inverve\_1DFFT\_Z}} ($(S_x/2+1) \times S_y$ rows, length $S_z$)

\item if (B$_z$ is free) \textbf{\textit{compactification\_Z+transpose}}

else \textbf{\textit{transpose}}

\item \textbf{\textit{inverse\_1DFFT\_Y}} ($(S_x/2+1) \times N_z$ rows, length $S_y$)

\item if (B$_y$ is free) \textbf{\textit{compactification\_Y+transpose}}

else \textbf{\textit{transpose}}

\item \textbf{\textit{inverse\_1DFFT\_X}} ($N_y \times N_z$ rows, length $S_x$) (C2R)

\item if (B$_x$ is free) \textbf{\textit{compactification\_X}}
\end{itemize}
}

In the implementation of the method we used the NVIDIA CUFFT1D library for evaluating batched 1D FFTs and the CUDA language
for transpositions and spread (compactification) operations. The method was tested with NVIDIA CUDA Toolkit version 4.1 on a NVIDIA 
Tesla series M2090 GPU using double precision. An OpenCL implementation of the method using custom 1D FFT GPU kernels is also in progress. 

\section{Results}

\begin{figure}[h]
 \centering
  \includegraphics[width=1.0\textwidth]{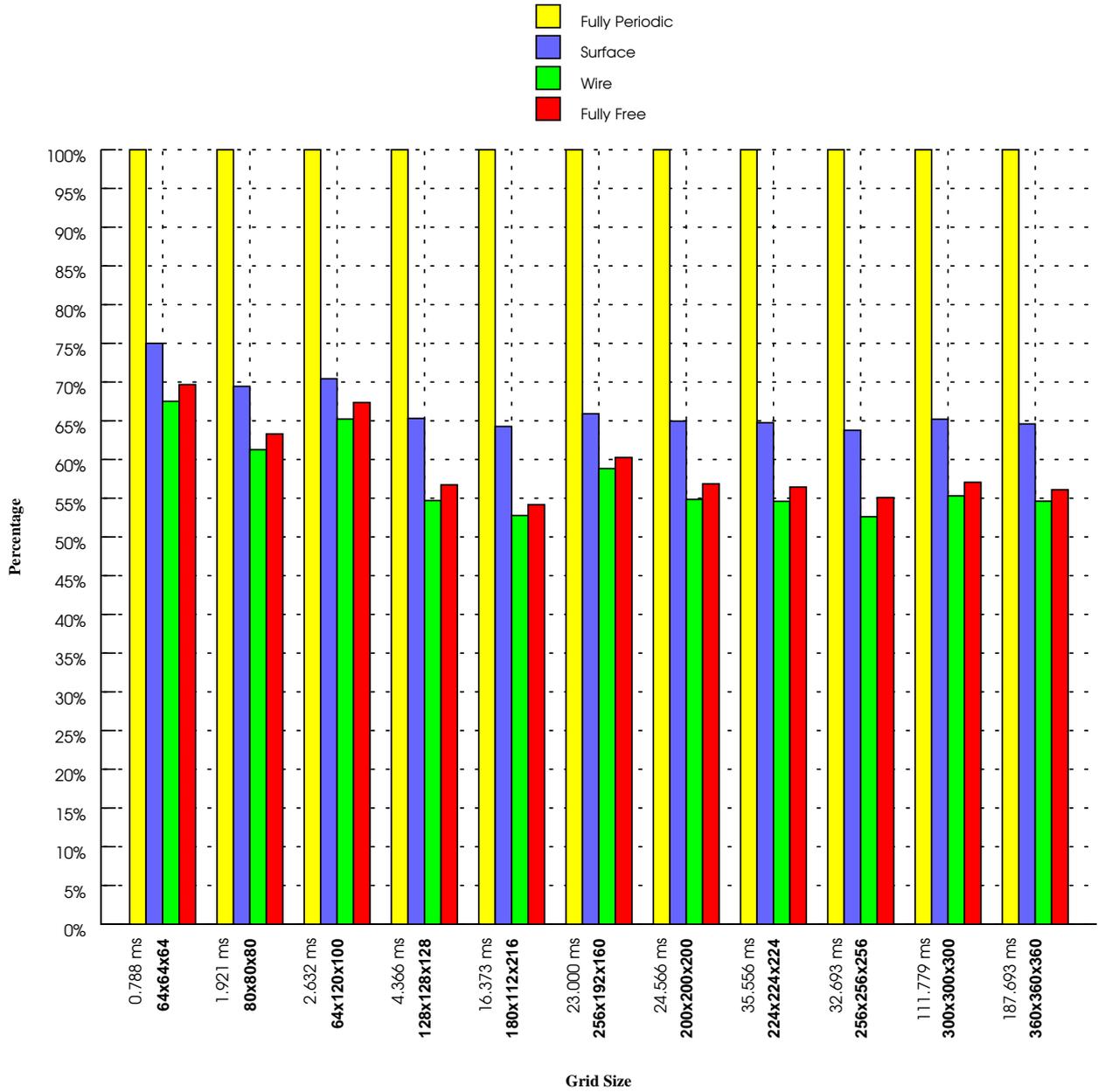}
 \caption{Comparison of the GPU computation times for different BCs. Fully free, wire and 
surface BCs computation times are shown as percentages of fully periodic BCs computation times which are given under 
corresponding column bars in milliseconds. The grid sizes given in the figure are FFT sizes ($S_x \times S_y \times S_z$) 
after the zero paddings in necessary dimensions.}
 \label{gpu}
\end{figure}

Benchmark calculations for fully periodic, surface, wire and fully free BCs are carried out by varying the grid size.
Computation times for the fully periodic case not
including the data transfers and computation times of the cases having one or more free boundary dimensions relative to the fully 
periodic case are given in Figure \ref{gpu} for different grid sizes. In the figure, grid sizes
are always given as the zero padded FFT sizes ($S_x \times S_y \times S_z$) instead of the input density sizes in order to keep FFT sizes
constant for different BCs, yielding a better comparison of the results. The computation time percentages for the free BCs case and the wire 
BCs case are close to the 
calculated percentage of $58\%$ for the amount of 1D FFTs except for the small grid dimensions. Small deviations from this value
are due to the spread, transposition and convolution kernel multiplication operations which become more important for small grid dimensions. 
For the surface BCs, again the computation time percentages are close to the calculated percentage of $66\%$ for the amount of 
1D FFTs except for the small grid dimensions and the variations can be explained with the same reasoning as the free and wire BCs 
cases.

Performance of the customized GPU Poisson solver in the presence of free dimensions is compared also with a CPU implementation of 
the same method where FFTW
library is used for batched 1D FFTs and CUDA kernels for spread and transposition operations are replaced with corresponding 
CPU functions in C programming language. Intel compiler version 11.1 is used in the compilation with default optimization level \linebreak (-O2)   
and {\it sse2} instructions are enabled in the compilation of the FFTW library. An Intel I7 930 2.8 GHz Quad Core processor
is used in these CPU calculations and the comparison is made against single core and also against 4 cores using OpenMP 
parallelization. GPU computation times with and without data transfer times are given for fully free, surface 
and wire BCs in Figure \ref{cpugpu1c} relative to single core CPU computation times and in Figure \ref{cpugpu4c} relative to 
4 cores CPU computation times. Grid sizes are given as the zero padded FFT sizes ($S_x \times S_y \times S_z$) also for these comparisons. 
For the single core case, computation time ratios between the CPU implementation  
and the GPU implementation vary in between $7.798$ and $24.389$ depending 
on the grid size and BCs. For the 4 cores CPU case 
the range of this ratio is between $2.769$ and $11.090$. When the transfer times of sending the input density to the device 
memory and getting the resulted potential from the device memory are also included in the comparison, the computation time 
ratios of the CPU case to the GPU case vary in between $2.879$ and $ 15.602$ for single core case and this ratio is between 
$1.154$ and $6.815$ for 4 cores CPU case.

In order to test the developed GPU Poisson solver in a real application, we implemented it in the BigDFT electronic
structure package and compared its performance with the CPU Poisson solver of the BigDFT for fully free BCs which is faster
than the CPU implementation using FFTW for fully free BCs since it takes advantage of merging transpose and kernel multiplication 
operations with customized 1D FFTs. These comparison results are given in Figure \ref{bigdft1c} and
Figure \ref{bigdft4c} for single core CPU and 4 cores CPU respectively. Including data transfer times, for the single 
core case computation time ratios 
between the CPU implementation and the GPU implementation vary in between $4.967$ and $14.922$ depending 
on the grid size. For the 4 cores CPU case the range of this ratio is between $1.610$ and $4.189$.

\section{Conclusions}

We have developed a customized GPU Poisson solver which is generally applicable to homogeneous and mixed BCs. 
Benchmark calculations show significant reduction of computation times compared to the fully periodic BCs case 
for the cases having one or more free BCs. The largest reduction of computation times are observed in the case of 
wire BCs where the periodic dimension is chosen as the first dimension of the forward 3D FFT. The reductions of 
computation times compared to the fully periodic case are in consistence with the reduction of total number of 1D FFT 
operations and these reductions would not be observed in the direct application of a 3D FFT library
for these cases with one or more free BC dimensions. Instead, in such a direct application, the computation times would 
slightly increase compared to the fully periodic case because of the additional zero padding spread operations and 
corresponding compactifications. However, the 
developers of various 3D FFT libraries may benefit from the techniques described in this work to increase the
performance of their libraries for such zero padded input data.

\begin{figure}[h]
 \centering
  \includegraphics[width=1.0\textwidth]{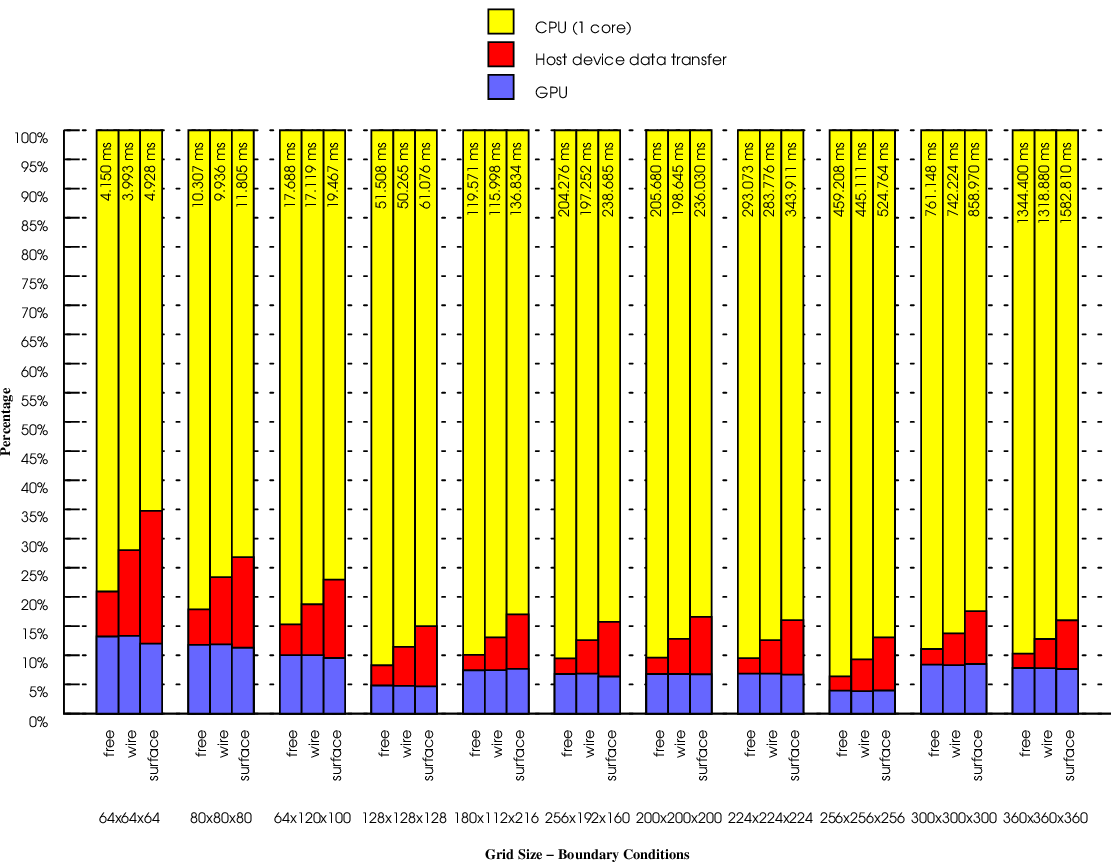}
 \caption{Comparison of the GPU computation times with FFTW implementation single core CPU
computation times. GPU and GPU + data transfer times are shown as percentages of single core CPU computation times for different 
grid sizes for fully free, wire and surface BCs. Single core CPU computation times are given on each column bar in milliseconds.
The grid sizes given in the figure are FFT sizes ($S_x \times S_y \times S_z$) after the zero paddings in necessary dimensions.}
 \label{cpugpu1c}
\end{figure}

\begin{figure}[h]
 \centering
  \includegraphics[width=1.0\textwidth]{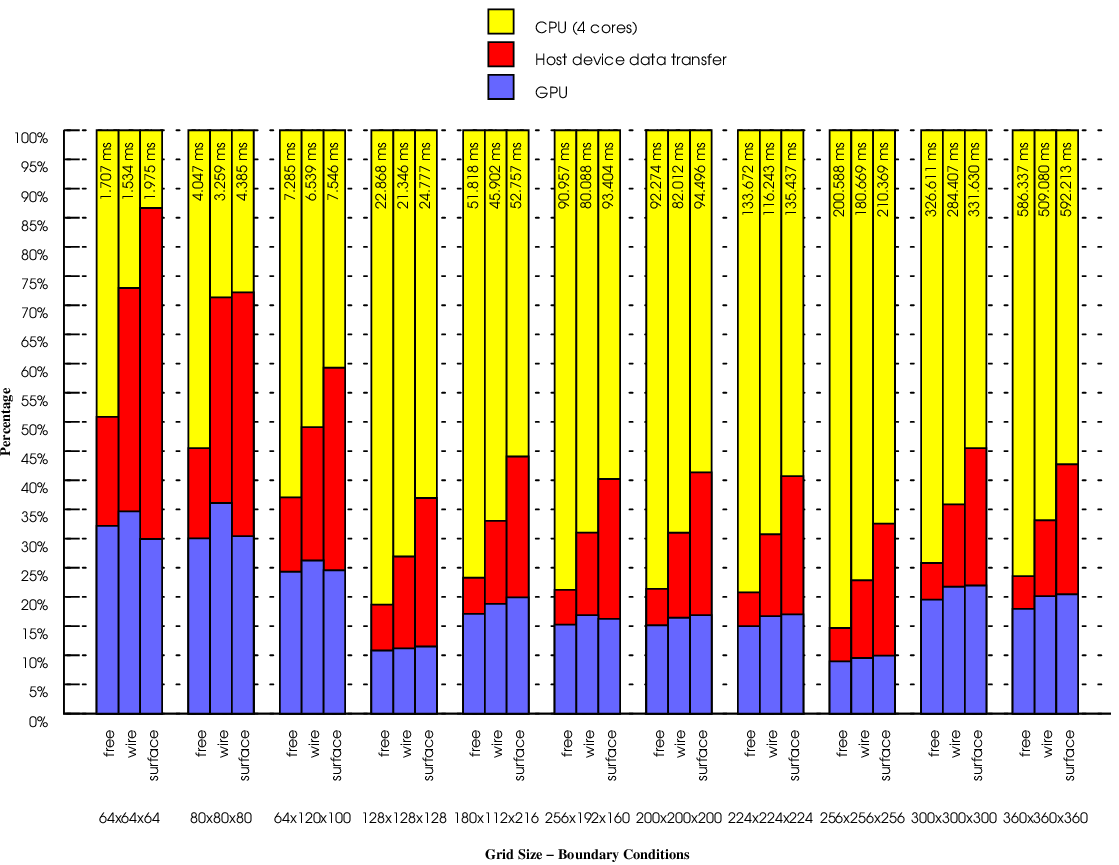}
 \caption{Comparison of the GPU computation times with FFTW implementation 4 cores CPU 
computation times. GPU and GPU + data transfer times are shown as percentages of 4 cores CPU computation times for different 
gird sizes for fully free, wire and surface BCs. 4 cores CPU computation times are given on each column bar in milliseconds.
The grid sizes given in the figure are FFT sizes ($S_x \times S_y \times S_z$) after the zero paddings in necessary dimensions.}
 \label{cpugpu4c}
\end{figure}

Comparison with CPU computations shows that the customized GPU Poisson solver provides up to $\sim$24.4 times faster computations
compared to single core CPU computation when data transfer times are not considered. 
With the advent of new hardware which merges CPU and GPU on the same chip and with new technologies like NVIDIA GPUDirect \cite{GPUDirect} 
which allows calling MPI subroutines between GPU memories, transfers to and from the GPU will be required less frequently and it 
will be possible to run a large fraction of sophisticated software packages on the GPU. 
Also, when multiple solution of the Poisson's equation from different input densities are required (as it is the case, for example, 
in Quantum Chemistry or electronic structure calculations), the transfer time for a given density can be overlapped by the 
calculation of the previous one \cite{Genovese.Ospici, Genovese.Videau}. For these reasons we give also the speedups without the transfer times.
Even when the transfer times are included,  the customized GPU Poisson solver provides up to $\sim$15.6 times 
faster computations compared to single core CPU. This means that the developed GPU Poisson solver can be used in real
applications including free BCs to accelerate the Poisson solver part of the code. Figures \ref{cpugpu1c} and \ref{cpugpu4c} 
show that the performance gain from GPU computation reduces for small sizes and it is maximal for large power of two
sizes. Increase of the data transfer to computation time ratios with increasing number of periodic dimensions is also
evident in these Figures. Therefore, when data transfer times are also considered, the amount of acceleration
due to GPU usage increases as the number of free dimensions increases. 

\begin{figure}[h]
 \centering
  \includegraphics[width=0.70\textwidth]{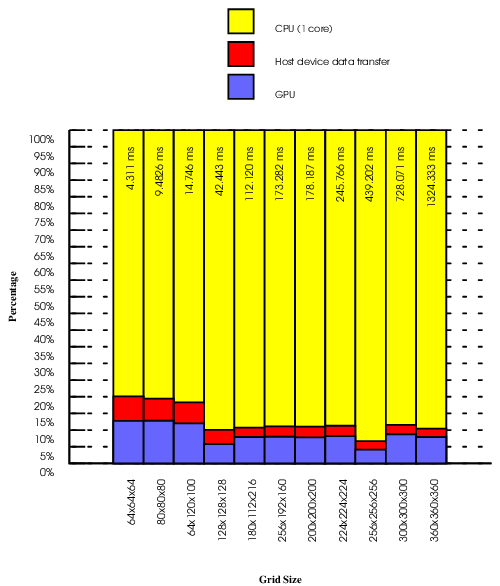}
 \caption{Comparison of the GPU computation times with BigDFT Poisson solver single core CPU
computation times for fully free BCs. GPU and GPU + data transfer times are shown as percentages of single core CPU computation 
times for different grid sizes. Single core CPU computation times are given on each column bar in milliseconds. The grid sizes 
given in the figure are FFT sizes ($S_x \times S_y \times S_z$) after the zero paddings in all dimensions.}
 \label{bigdft1c}
\end{figure}

\begin{figure}[h]
 \centering
  \includegraphics[width=0.70\textwidth]{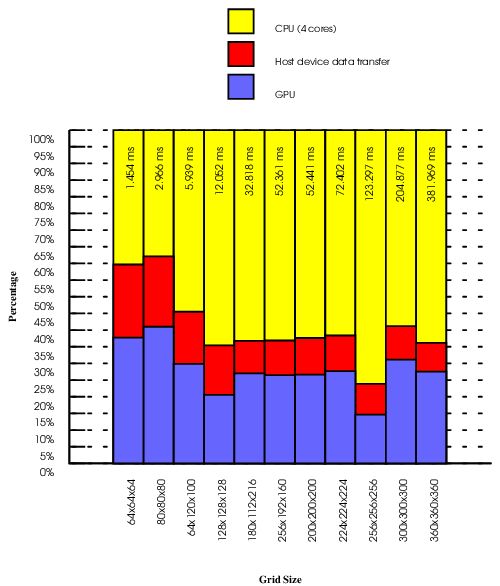}
 \caption{Comparison of the GPU computation times with BigDFT Poisson solver 4 cores CPU 
computation times for fully free BCs. GPU and GPU + data transfer times are shown as percentages of 4 core CPU computation 
times for different grid sizes. 4 cores CPU computation times are given on each column bar in milliseconds. The grid sizes given 
in the figure are FFT sizes ($S_x \times S_y \times S_z$) after the zero paddings in all dimensions.}
 \label{bigdft4c}
\end{figure}

The new developments are implemented in the BigDFT Electronic Structure Package 
version 1.7 and the Poisson solver part of the code is accelerated as shown in Figure 
\ref{bigdft1c} and Figure \ref{bigdft4c} for fully free BCs calculations. Up to $\sim$14.9 times acceleration compared to single
core CPU is observed
in these tests including data transfer times. For surface and wire BCs, the comparison was not carried out 
since BigDFT code does not consider the optimal order of periodic and free dimensions for these mixed BCs cases at the moment. 
The source code of the BigDFT package can be accessed from the BigDFT project web page \cite{BigDFT} under GPL License agreement. 

The convolution operation which is the main part of the Poisson solver discussed in this work is not only required in the 
context of solving Poisson's equation, but also in many other contexts.
In plane wave density functional programs, the application of the local  potential onto the wave function is for instance a 
convolution of the same type as the convolution in the solution of Poisson's equation with free BCs. 
Such programs have also been ported to GPU's \cite{Wang.Jia} but the possible speedup that could be obtained in this context 
by a customized FFT has not been exploited so far. 
\\
\\
\noindent {\bf Acknowledgements:} We thank Peter Messmer from NVIDIA Corp. for his suggestions and valuable discussions about the
subject.

\bibliographystyle{model1-num-names}
\bibliography{psolver_gpu}

\onecolumn

\end{document}